\documentclass[aps,preprintnumbers,eqsecnum,amsmath,amssymb,showpacs,nofootinbib]{revtex4}
\usepackage{graphicx}
\usepackage{dcolumn}
\usepackage{bm}
\usepackage{epsfig}

\begin{document}
\title{Pion form factor at spacelike momentum transfers from local-duality QCD sum rule}  
\author{Victor Braguta$^a$, Wolfgang Lucha$^{b}$, and Dmitri Melikhov$^{b,c}$}
\affiliation{
$^a$ Institute for High Energy Physics, 142284, Protvino, Russia\\
$^b$ Institute for High Energy Physics, Austrian Academy of Sciences, Nikolsdorfergasse 18, A-1050, Vienna, Austria\\
$^c$ Nuclear Physics Institute, Moscow State University, 119991, Moscow, Russia}
\date{\today}
\begin{abstract}
We study the pion form factor in a broad range of spacelike momentum transfers 
within the local-duality version of QCD sum rules.   
We make use of the recently calculated two-loop double spectral density 
of the $\langle AVA\rangle$ correlator 
including $O(1)$ and $O(\alpha_s)$ terms, which allows us to give predictions 
for the pion form factor and to study the interplay between the 
nonperturbative and perturbative contributions to the pion form factor without 
any reference to the pion distribution amplitude. 
Our results demonstrate the dominance of the 
nonperturbative contribution to the form factor up to relatively large values of the momentum 
transfer: namely, the nonperturbative $O(1)$ term, which provides the $1/Q^4$ power correction, gives more
than half of the pion form factor in the region $Q^2 \le 20$ GeV$^2$. 
\end{abstract}
\maketitle

\section{Introduction}
The study of the interplay between perturbative and nonperturbative physics in exclusive processes, 
and, in particular, in the pion form factor which we discuss in this paper, has a long history. 
At asymptotically large $Q^2$, the pQCD factorization formula 
\cite{pqcd} gives the pion form factor in terms of the scale-dependent 
pion distribution amplitude (DA) 
of leading twist $\phi_\pi(u,Q^2)$:  
\begin{eqnarray}
\label{fact}
F_\pi(Q^2)=\frac{8\pi \alpha_s(Q^2) f_\pi^2}{9Q^2} 
\left|\int\limits_0^1 du\frac{\phi_\pi(u,Q^2)}{u}\right|^2. 
\end{eqnarray}
The DA, as obtained from the pQCD evolution equation, 
has the form   
\begin{eqnarray}
\label{phiasym}
\phi_\pi(u,Q^2\to\infty)=6u(1-u).
\end{eqnarray}
Respectively, at asymptotically large $Q^2$ a direct prediction of pQCD reads \cite{pqcd,jackson}
\begin{eqnarray}
\label{ffass}
Q^2 F_\pi(Q^2)=8\pi \alpha_s(Q^2)f_\pi^2. 
\end{eqnarray}
Subleading logarithmic and power corrections to this formula should be taken into account 
at large but finite $Q^2$. This is, however, a very difficult task. 
There are two competitive scenarios for the pion form factor at intermediate momentum transfers:  

The first scenario (A) is based on the assumption that power corrections are negligible 
in the region $Q^2\ge 3$--$5$ GeV$^2$. The form factor is then 
given by the pQCD factorization formula (\ref{fact}) with the pion
distribution amplitude at low normalization scale, which in this scenario turns out to have 
a double-humped ``camel'' shape with an enhanced end-point region \cite{cz}, very different 
from its asymptotic form (\ref{phiasym}). 
This scenario, complemented by the analysis of Sudakov double logarithms \cite{sterman}, provides 
the basis for the perturbative QCD approach to form factors at intermediate momentum transfers. 

In the second scenario (B), which we consider to be more realistic, 
the form factor is dominated by the nonperturbative contributions up to 
rather high values of $Q^2$, 
with the perturbative contribution remaining relatively small \cite{isgur,nesterenko}.  
The end-point behaviour of the DA at a low normalization scale is then similar to that of the 
asymptotic DA (\ref{phiasym}). 
This scenario is supported by the fact that the soft contribution to the form factor 
alone can reproduce the pion form factor to a good accuracy for $Q^2$ up to several 
GeV$^2$ \cite{nesterenko,ioffe,bakulev,braun1,simula,m}. 

In \cite{anis}, making use of the constituent quark picture, and in \cite{braun2}, 
within light-cone sum rules, 
the pion form factor was analyzed by taking into account the nonperturbative $O(1)$ contribution 
and the radiative $O(\alpha_s)$ corrections. The form factor at intermediate $Q^2$ 
turned out to be sensitive to the details of the pion wave function --- the Bethe-Salpeter 
wave function in \cite{anis} and the pion light-cone 
DA in \cite{braun2}. Scenario B was favoured by these results.  

Unfortunately, the data on the pion form factor for $Q^2>2$ GeV$^2$ are not sufficiently precise,  
leaving room for speculations about the details of the pion DA at low normalization scale 
and, respectively, on the relative weights of the soft and the hard contributions to the pion 
form factor. 

Therefore, it seems interesting to address the problem without a direct reference to the pion DA. 
The local-duality version of three-point QCD sum rules \cite{nesterenko} provides this opportunity.  

The local-duality sum rule is the Borel sum rule in the limit of an infinitely large Borel parameter.   
For the relevant choice of the pion interpolating current, 
the condensate contributions to the correlator vanish in this limit and the pion 
observables are given by dispersion integrals via the spectral densities of purely perturbative QCD diagrams. 
The integration region in the dispersion integrals is restricted to the pion ``duality interval''. 

In this Letter we apply a local-duality sum rule to the pion form factor,  
making use of the recently calculated two-loop double spectral density of 
the pion form factor for massless quarks \cite{braguta2,braguta}.
Such an approach has the following
attractive features: 
(i) it is applicable in a broad range of momentum transfers starting from low 
to asymptotically large values, and 
(ii) it does not refer to the pion distribution amplitude. 
Therefore, it allows us to study in a relatively model-independent way the interplay 
between perturbative and nonperturbative dynamics in the pion from factor.


\section{Sum rule}
We shall consider the pion form factor in the chiral limit of massless quarks and 
a massless pion. Let us recall well-known results for Borel sum rules:   
The sum rule for the pion decay constant is obtained from the 
OPE for the two-point function and reads \cite{svz} 
\begin{eqnarray}
\label{fpi}
f_\pi^2 = \frac{1}{\pi}\int_0^{s_0}ds
\exp\left(-{s}/{M^2}\right)\rho(s)
+
\frac{\langle\alpha_s G^2\rangle}{12 \pi M^2}
+
\frac{176 \pi\alpha_s \langle \bar qq \rangle^2}{81M^4}+\cdots,  
\end{eqnarray}
where $\rho(s)=\frac{1}{4\pi}\left(1+\frac{\alpha_s}{\pi}\right)+O(\alpha_s^2)$ 
is the perturbative spectral density. 

The Borel sum rule for the pion form factor is obtained from the OPE for the three-point function 
and reads \cite{nesterenko,ioffe} 
\begin{eqnarray}
\label{ff}
f_\pi^2 F_{\pi} (Q^2)= 
\Gamma(Q^2,M^2,M^2|s_0)
+\frac {\langle \alpha_sG^2\rangle}{24\pi M^2}
+\frac {4\pi\alpha_s \langle \bar qq \rangle^2}{81 M^4 }\left(13+\frac{Q^2}{M^2}\right). 
\end{eqnarray}
Here, $\Gamma(Q^2,M^2,M^2|s_0)$ is the perturbative contribution, which is obtained by the following procedure:
One calculates the double Borel transform of the 
$\langle AVA\rangle$ correlator 
\begin{eqnarray}
\Gamma(Q^2,M_1^2,M_2^2)=
\frac{1}{\pi^2}\int ds_1 ds_2 \exp(-s_1/M_1^2) \exp(-s_2/M_2^2)
\left[\Delta^{(0)}(Q^2,s_1, s_2)+ \alpha_s \Delta^{(1)}(Q^2,s_1, s_2)\right],  
\label{borelcor}
\end{eqnarray}
and restricts the integration in the $s_1$--$s_2$ plane to the pion duality region. 
One then sets $M_1^2\to 2M^2$, $M_2^2\to 2M^2$ and compares the two- and three-point sum 
rules for the same values of the Borel parameter $M^2$.\footnote{Notice
that such a procedure of comparing two- and three-point sum rules finds a natural physical explanation 
within the correspondence between sum rules and the constituent quark picture observed in \cite{ms}.} 
The function $\Delta^{(0)}(Q^2,s_1, s_2)$ is well-known, whereas $\Delta^{(1)}(Q^2,s_1, s_2)$ was calculated 
only recently \cite{braguta} for the case of massless quarks.\footnote{
Another case in which the radiative corrections to the double spectral density of the three-point function 
have been calculated is the case of one massless and one infinitely heavy quark \cite{colangelo}.} 
The explicit expressions can be found in \cite{braguta}.

The local-duality (LD) sum rules \cite{nesterenko,bakulev,radpol} correspond to the limit $M\to\infty$. 
A remarkable feature of this limit is the vanishing of the condensate contributions to the sum rules 
for the pion form factor and the decay constant. Assuming the duality 
region in the $s_1$--$s_2$ plane to be a square of side $s_0$, 
and denoting the duality interval in the sum rule for the decay constant by 
$\bar s_0$, we obtain to $\alpha_s$ accuracy 
\begin{eqnarray}
\label{ldff}
f_\pi^2 F_{\pi} (Q^2) &=& 
\frac{1}{\pi^2}\int_0^{s_0}ds_1\int_0^{s_0}ds_2 
\left[\Delta^{(0)} (s_1, s_2, Q^2)+ \alpha_s \Delta^{(1)} (s_1, s_2, Q^2)+O(\alpha_s^2)\right], 
\\ 
\label{ldfpi}
f_\pi^2 &=& 
\frac{1}{\pi}\int_0^{\bar s_0}ds\left[\rho^{(0)}(s)+\alpha_s\rho^{(1)}(s)+O(\alpha_s^2)\right]=
\frac{\bar s_0}{4\pi^2}\left(1+\frac{\alpha_s}{\pi}\right)+O(\alpha_s^2). 
\end{eqnarray}
One should not be confused by the simplicity of these expressions: the complicated nonperturbative
dynamics is now hidden in the effective continuum thresholds $s_0$ and $\bar s_0$. Let us emphasize 
that the LD sum rules are predictive only if one knows, or fixes according to some criteria, 
the effective continuum thresholds. 

Some comments on the dispersion representations for $f_\pi^2 F_\pi$ and $f_\pi^2$ are in order: 

\noindent 1. 
Whereas the single dispersion representation for the decay constant $f_\pi^2$ is well-defined, the double
dispersion representation for $f_\pi^2F_\pi $, even in the LD limit, 
has at least two essential ambiguities: 

a. The shape of the duality region in the $s_1$--$s_2$ plane: 
the simplest choice is a square, but any other region symmetric under $s_1\leftrightarrow s_2$ may be 
also possible. 

b. Nothing forbids the upper boundary of the duality region $s_0$ from being $Q^2$-dependent, and 
additional assumptions to fix $s_0(Q^2)$ are necessary. 
Arguments in favour of choosing 
the parameters in two- and three-point sum rules constant and equal to each other were given in \cite{bakulev}.  
Let us see what happens if we choose the same constant value $\bar s_0=s_0$ in the sum rules
(\ref{ldff}) and (\ref{ldfpi}), and substitute the sum rule (\ref{ldfpi}) instead of $f_\pi^2$ 
into (\ref{ldff}): 
\begin{eqnarray}
\label{ldff2}
F_{\pi} (Q^2) &=&\frac{\displaystyle
\frac{1}{\pi^2}\int_0^{s_0}ds_1\int_0^{s_0}ds_2 
\left[\Delta^{(0)} (s_1, s_2, Q^2)+ \alpha_s \Delta^{(1)} (s_1, s_2, Q^2)\right]}
{\displaystyle\frac{1}{\pi}\int_0^{s_0}ds\left[\rho^{(0)}(s)+\alpha_s\rho^{(1)}(s)\right]}, 
\qquad s_0=\frac{4\pi^2 f_\pi^2}{1+{\alpha_s}/{\pi}}.
\end{eqnarray}

The LD form factor given by (\ref{ldff2}) has the following interesting properties \cite{lucha}: 

(i) It satisfies the normalization condition $F_\pi(Q^2=0)=1$ due to the vector 
Ward identity which relates the spectral density of the self-energy diagram and the double 
spectral density of the triangle diagram at zero momentum transfer:  
\begin{eqnarray}
\label{wi}
\lim_{Q^2\to 0}\Delta^{(i)}(s_1,s_2,Q^2)=\pi \rho^{(i)}(s_1)\delta(s_1-s_2),
\qquad \rho^{(0)}(s)=\frac{1}{4\pi},
\qquad \rho^{(1)}(s)=\frac{1}{4\pi^2}. 
\end{eqnarray}
Clearly, for consistency one should then take into account the radiative corrections 
to the same order in two- and three-point correlators. 

(ii) Making use of the explicit expressions for $\Delta^{(i)}$, one obtains 
\begin{eqnarray}
\lim_{Q^2\to \infty}\Delta^{(0)}(s_1,s_2,Q^2)&=&\frac{3(s_1+s_2)}{2 Q^4},\\
\qquad \lim_{Q^2\to \infty}\Delta^{(1)}(s_1,s_2,Q^2)&=&\frac{1}{2\pi Q^2}. \label{d1}  
\end{eqnarray}
Substituting these expressions into (\ref{ldff2}), one finds at large $Q^2$:
\begin{eqnarray}
\label{norunning}
F_\pi(Q^2)=\frac{8\pi f_\pi^2\alpha_s}{Q^2}+\frac{96\pi^4 f_\pi^4}{Q^4}
+O\left(\alpha_sf_\pi^4/Q^4\right)+O\left(\alpha_s^2\right).
\end{eqnarray}
We find quite remarkable that the exact $\Delta^{(1)}$ leads to the correct pQCD  
(up to the running of $\alpha_s$) large-$Q^2$ asymptotics of the pion form factor 
obtained from the LD sum rule (\ref{ldff2}). Let us explain this important point: 
Whereas, e.g., the normalization of the pion form factor (\ref{ldff2}) at $Q^2=0$ is the 
consequence of the Ward identity, 
we do not see any {\it rigorous} condition which would guarantee the correct large-$Q^2$ behaviour when using 
the same value of the pion duality intervals in two- and three-point correlators.    
We find this to be a strong argument in favour of the universality of the pion duality interval.  

\begin{figure}[t]
\begin{tabular}{cc}
\includegraphics[width=8cm]{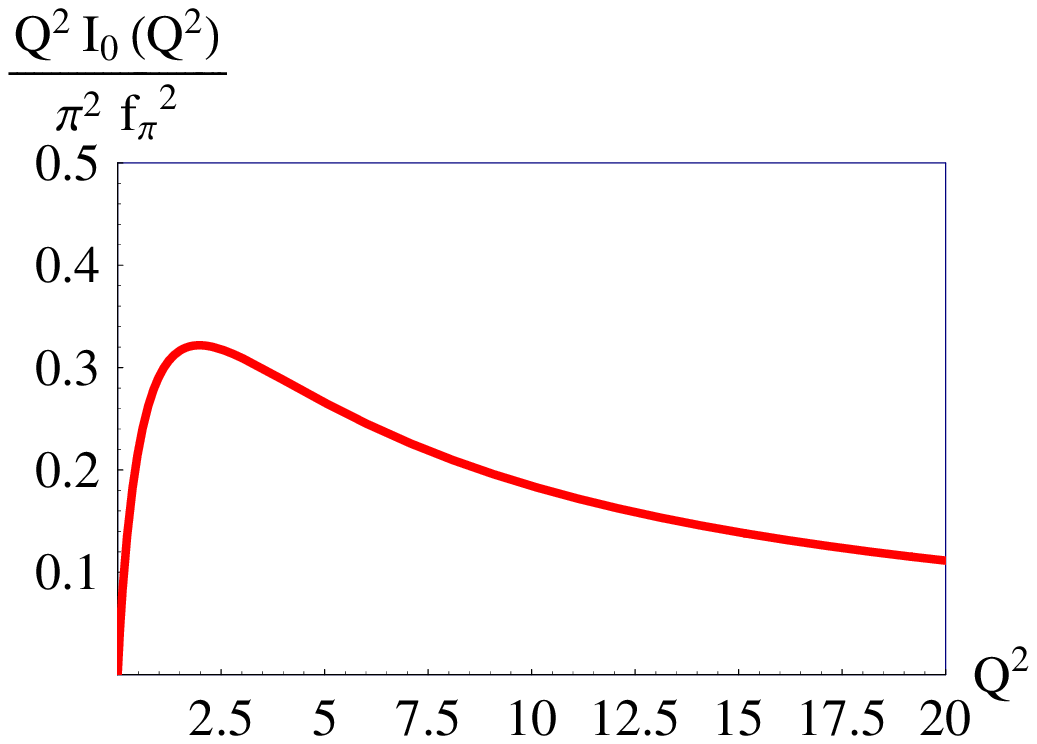}&\includegraphics[width=8cm]{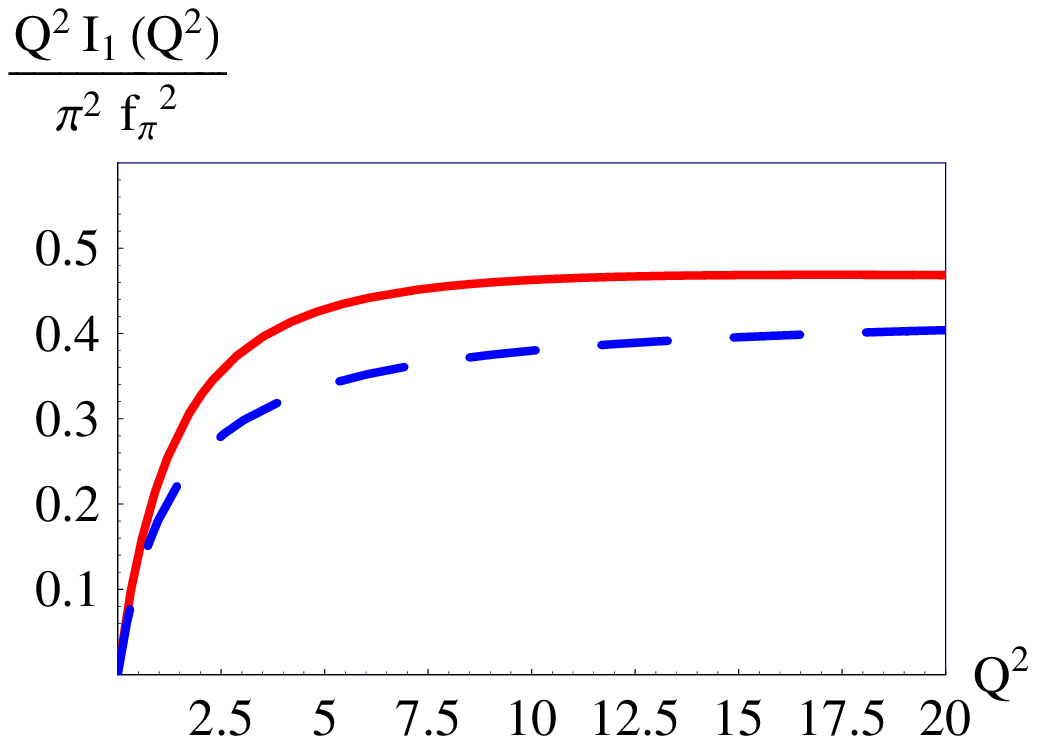}
\end{tabular}
\caption{\label{Fig:0} $\frac{Q^2}{\pi^2f_\pi^2}I_0(Q^2,s_0)$ (a) and 
$\frac{Q^2}{\pi^2f_\pi^2}I_1(Q^2,s_0)$ (b) 
for $s_0=4\pi^2f_\pi^2$. 
Solid (red) line: exact result, dashed (blue) line: Ansatz (\ref{interpol}).}
\end{figure}

(iii)
The $O(1)$ contribution, shown in Fig.~\ref{Fig:0}a, was calculated in \cite{nesterenko}: 
\begin{eqnarray}
\label{i0}
I_0(Q^2,s_0)=\int_{0}^{s_0} ds_1 \int_{0}^{s_0} ds_2\, \Delta^{(0)}(s_1,s_2,Q^2)=
\frac{s_0}{4}\left(1-\frac{1+6s_0/Q^2}{(1+4s_0/Q^2)^{3/2}}\right). 
\end{eqnarray}
The  explicit expression for the $O(\alpha_s)$ contribution 
\begin{eqnarray}
\label{i1}
I_1(Q^2,s_0)=\int_{0}^{s_0} ds_1 \int_{0}^{s_0} ds_2 \,\Delta^{(1)}(s_1,s_2,Q^2)
\end{eqnarray}
was obtained only recently \cite{braguta}. Before that, it was proposed to use 
instead of the unknown integral the simplest Ansatz \cite{radpol}
\begin{eqnarray}
\label{interpol}
I_1(Q^2,s_0)\rightarrow \frac{s_0}{4\pi}\frac{1}{1+Q^2/2s_0},  
\end{eqnarray}
which reproduces the value of the integral at $Q^2=0$, fixed by the Ward identity,  
and its asymptotic behaviour according to (\ref{d1}). Fig.~\ref{Fig:0}b compares the 
formula (\ref{interpol}) with the result of the exact calculation: as one can see, 
the proposed formula underestimates the exact $I^{(1)}$ by more than 20\% in the broad 
range of practical relevance $Q^2=$1--30 GeV$^2$. 

\newpage
\noindent 2. There are obvious problems with the application of this sum rule at small $Q^2\le 1$ GeV$^2$: 

First, the OPE for the three-point correlator was obtained in the region where 
all three external variables $|p_1^2|$, $|p_2^2|$, and $Q^2$ are large. 
Therefore, the sum rule cannot be
directly applied at small $Q^2$, although the expression (\ref{ldff2}) leads to the correct normalization 
of the form factor. Additional contributions appear at small $Q^2$ \cite{radsmallq2} 
which prevent from giving unambiguous predictions in this region. 
Of course, allowing for a $Q^2$-dependent value $s_0(Q^2)$ in the sum rule (\ref{ldff}), 
we can formally extend the formula also to lower $Q^2$ and apply it starting from $Q^2=0$, 
but as we have noticed above, in this case the sum rule loses its predictivity.{ 
A technical indication that the LD sum rule (\ref{ldff2}) cannot be applied at very small $Q^2$ 
is the presence of the terms $\sim \sqrt{Q^2}$ [see (\ref{i0})]  
leading to an infinite value of the pion radius.}

Second, the spectral density $\Delta(s_1,s_2,Q^2)$ contains the terms $O(1)$ and 
$O(\alpha_s)$, whereas higher powers are unknown. Since the 
coupling constant $\alpha_s$ is not small in the soft region,
our spectral density is not sufficient for application to the form factor 
at $Q^2\le 1$ GeV$^2$.

\vspace{.2cm}
\noindent 3.
In order to apply the obtained formulas for large $Q^2$, higher-order radiative corrections, 
leading to the running of $\alpha_s$, should be taken into account. Such an accuracy is beyond our 
two-loop calculation; nevertheless, a self-consistent expression for 
the form factor applicable for all $Q^2>0$ may be written as 
\begin{eqnarray}
\label{ldff3}
F_{\pi} (Q^2) &=&\frac{1}{f_\pi^2 \pi^2}
\int_0^{s_0(Q^2)}ds_1\int_0^{s_0(Q^2)}ds_2 
\left[\Delta^{(0)} (s_1, s_2, Q^2)+ \alpha_s(Q^{*2})\Delta^{(1)} (s_1, s_2, Q^2)\right], 
\end{eqnarray}
where the scale $Q^{*2}$ in the argument of $\alpha_s$ is related to $Q^2$ 
(see the discussion in \cite{bakulev2}) and $s_0(Q^2)$ satisfies the boundary conditions\footnote{
In \cite{radpol} it was argued that $s_0$ is the relevant scale of $\alpha_s$ in the LD sum rules 
for the decay constant 
and for the form factor at $Q^2=0$. For our discussion this subtlety is irrelavant so we 
somewhat symbolically use the notation $\alpha_s(0)$. } 
\begin{eqnarray}
\label{boundary}
s_0(Q^2=0)=\frac{4\pi^2f_\pi^2}{1+\alpha_s(0)/\pi},\qquad s_0(Q^2=\infty)={4\pi^2f_\pi^2}. 
\end{eqnarray}
If the effective threshold $s_0(Q^2)$ satisfies these relations, the form factor 
is normalized to $F_\pi(0)=1$ and reproduces the pQCD asymptotic behaviour at $Q^2\to\infty$. 

In the following, we shall set the scale $Q^{*2}=Q^2$: 
in the region $Q^2\ge 1$ GeV$^2$, $\alpha_s(Q^2)$ is a slowly varying function of $Q^2$ 
(Fig.~\ref{Fig:1}a); 
therefore, the precise setting of the scale makes very little difference. 
We shall use the following appealing parametrization of $s_0(Q^2)$, 
obviously satisfying (\ref{boundary}):  
\begin{eqnarray}
\label{threshold3}
s_0(Q^2)=\frac{4\pi^2f_\pi^2}{1+{\alpha_s(Q^2)}/{\pi}}. 
\end{eqnarray}
Before turning to the numerical analysis, we would like to draw the reader's attention to the following 
observation: An essential feature of the form factor obtained from the three-point sum rule is the full 
cancellation of the double logarithmic terms. The proof of this general property 
of the color-neutral three-point Green functions in QCD can be found in \cite{cz} (see also \cite{tiktopoulos});  
the cancellation of double logs was checked in explicit two-loop calculations for various 
quark currents in \cite{braguta,braguta2}. 
In contrast to this result, the pion form factor obtained from the light-cone sum rule 
contains double log terms \cite{braun2}. This discrepancy requires a clarification. 
Presumably, higher-twist contributions, which were not taken into account in \cite{braun2}, but which are in general not 
suppressed compared to the lower-twist contributions \cite{lms_lc} play a crucial role here. 


\section{Numerical results}
\begin{figure}[h]
\begin{tabular}{cc}
\includegraphics[width=8cm]{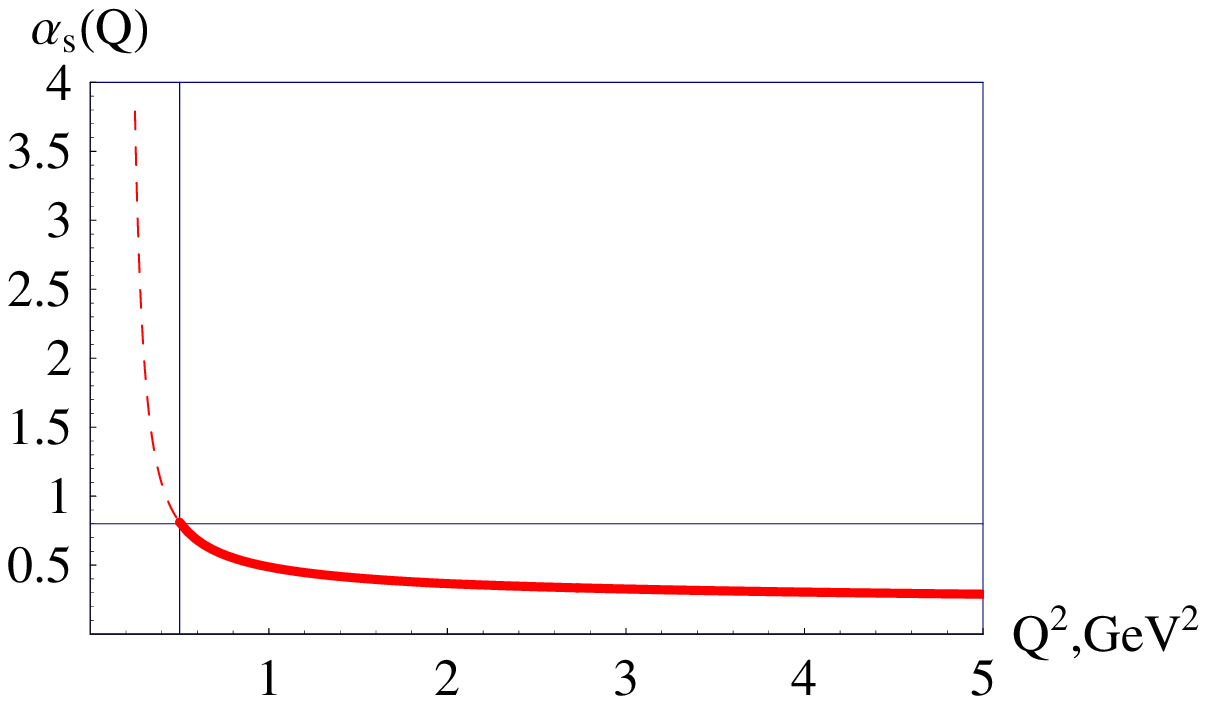}&\includegraphics[width=8cm]{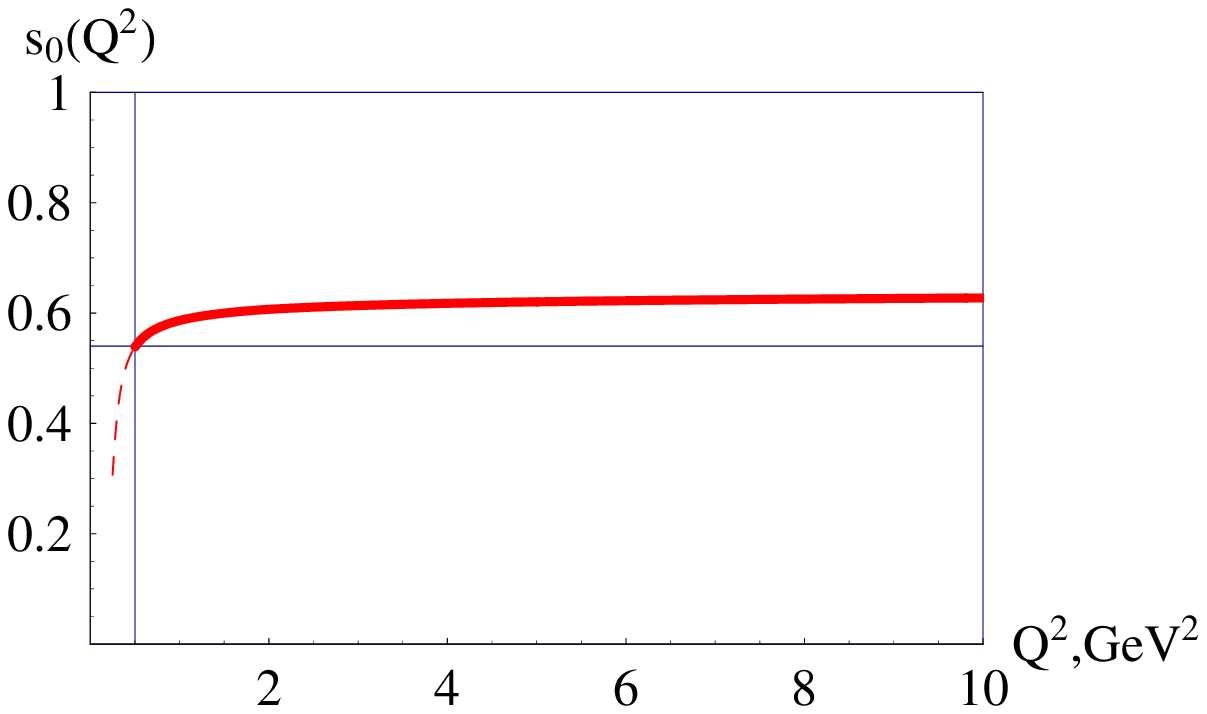}
\end{tabular}
\caption{\label{Fig:1}The perturbative $\alpha_s(Q)$ (a) and the corresponding effective threshold $s_0(Q^2)$ 
(b) given by (\ref{boundary}). Dashed lines show these quantities outside our working region.}
\end{figure}

For numerical estimates, we make use of the three-loop running $\alpha_s(Q^2)$ (Fig.~\ref{Fig:1}a). 
%
The corresponding $s_0(Q^2)$ is shown in Fig.~\ref{Fig:1}b. Notice that it is a slowly varying function 
in the region $Q^2\ge 1$ GeV$^2$, where we apply the LD sum rule. 

\begin{figure}[h]
\begin{tabular}{cc}
\includegraphics[width=8cm]{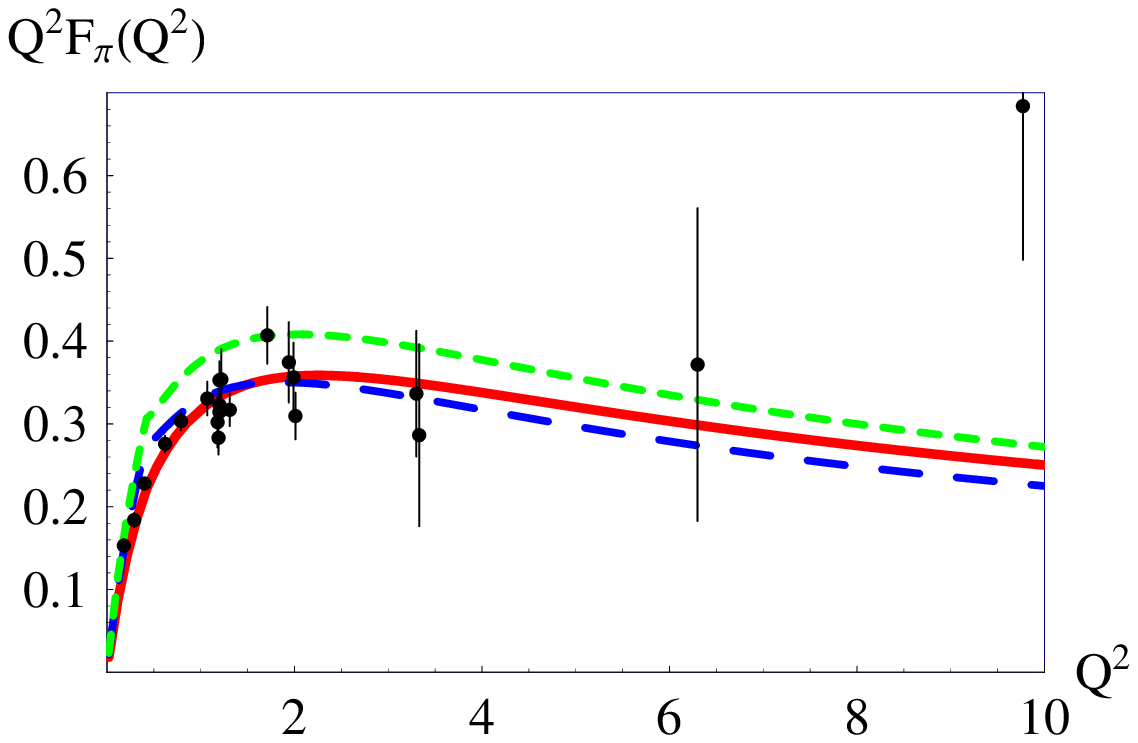}&\includegraphics[width=8cm]{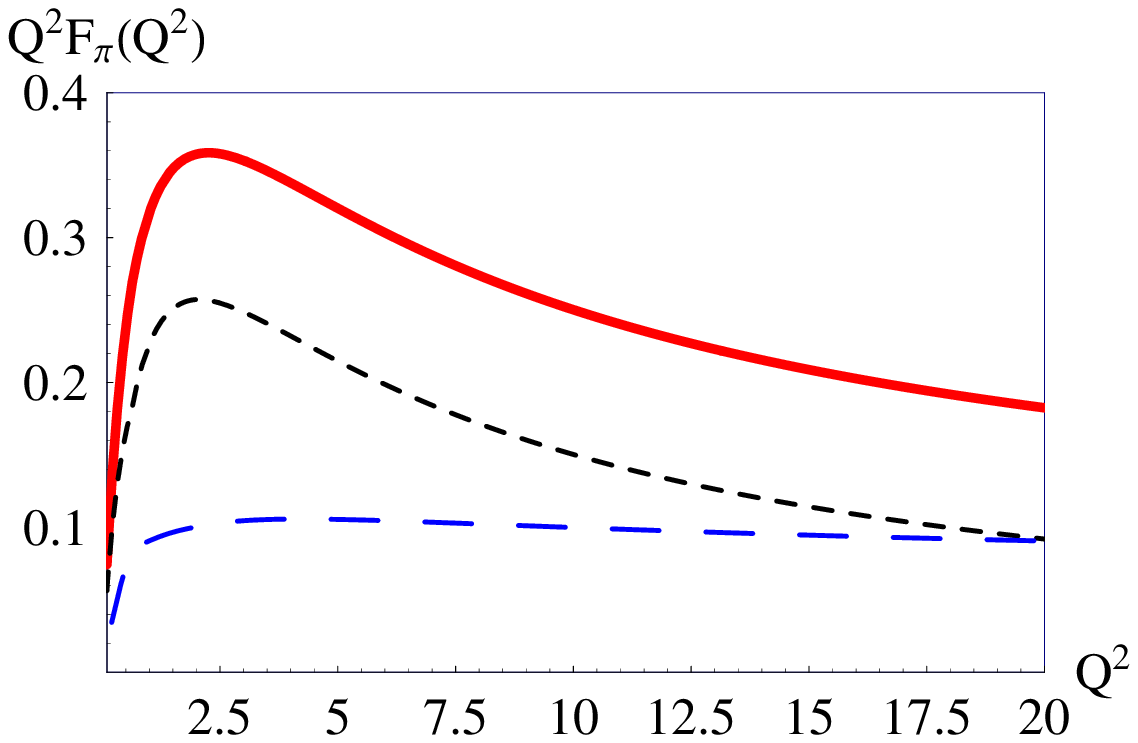}
\end{tabular}
\caption{\label{Fig:2}The pion form factor at $Q^2\ge 0.5$ GeV$^2$. 
Experimental data from \cite{data_largeQ2}. 
Solid (red) line: the result of the calculation according to (\ref{ldff3}). 
(a) Short-dashed (green) line: the form factor obtained with constant $s_0=0.65$ GeV$^2$; 
long-dashed (blue) line: $s_0=0.6$ GeV$^2$. (b) Short-dashed (black) line: 
the $O(1)$ contribution, long-dashed (blue) line: the $O(\alpha_s)$ contribution.}
\end{figure}

The pion form factor is shown in Fig.~\ref{Fig:2}a. The $O(1)$ and $O(\alpha_s)$ terms, separately, are given
in Fig.~\ref{Fig:2}b. It should be noticed that the $O(1)$ term providing the $1/Q^4$ power correction 
at large $Q^2$, dominates the form factor at low $Q^2$, and still gives 50\% at $Q^2=20$ GeV$^2$. The
$O(\alpha_s)$ term gives more than 80\% of the form factor only above $Q^2=100$ GeV$^2$. 
Such a pattern of the pion form factor behaviour has been conjectured many times in the literature; 
we now obtain this behaviour in an explicit calculation. The earlier analyses of the pion form factor 
in a broad range of momentum transfers \cite{anis,braun2,bakulev2,craig} are consistent with the 
results reported in Fig.~\ref{Fig:2} within about 20\% accuracy. 

\begin{figure}[ht]
\begin{tabular}{c}
\includegraphics[width=7cm]{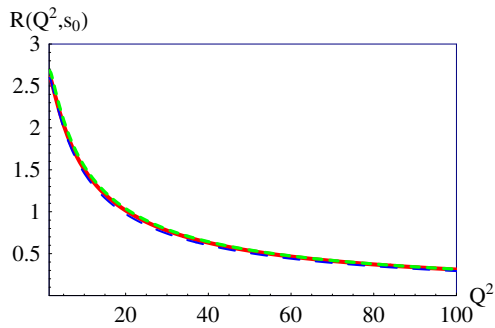}
\end{tabular}
\caption{\label{Fig:3}
The ratio of the $O(1)$ and $O(\alpha_s)$ contributions to the pion form factor
$R(Q^2,s_0)=\displaystyle{I_0(Q^2,s_0)}/{[\alpha_s(Q^2)I_1(Q^2,s_0)]}$. 
Solid (red) line: the result of the calculation with the effective continuum threshold $s_0(Q^2)$ 
(\ref{threshold3}), short-dashed (green) line: $s_0=0.65$ GeV$^2$, 
long-dashed (blue) line: $s_0=0.6$ GeV$^2$.
}
\end{figure}

Fig.~\ref{Fig:3} presents the ratio of the $O(1)$ and the $O(\alpha_s)$ contributions to the pion form
factor vs $Q^2$ for different models of the effective continuum thresholds. One can clearly see 
that the ratio is mainly determined by the corresponding double spectral densities 
$\Delta^{(0)}$ and $\Delta^{(1)}$, whereas its  
sensitivity to the effective continuum threshold is rather weak.


\section{Discussion and Conclusions}
We have presented the analysis of the pion form factor in a broad range of spacelike momentum 
transfers making use of the local-duality sum rule. 
This is the first analysis which takes into account both the leading order $O(1)$ contribution to the pion form factor 
and the recently calculated first-order $O(\alpha_s)$ radiative correction. 
These ingredients are crucial for the possibility to consider the form factor in a broad range of $Q^2$ and to study   
the transition from the nonperturbative to the perturbative region. 

Let us summarize the essential ingredients, the uncertainties, and the lessons to be learnt from our analysis:  
\begin{itemize}
\item
{\it The double spectral density of the spectral representation for the form factor}: 
We have good control over the spectral density --- we have included the exact 
$O(1)$ and $O(\alpha_s)$ terms, and omit the (unknown) $O(\alpha_s^2)$ terms, 
which are expected to contribute less than 10\% at $Q^2>$ 1 GeV$^2$. 
[The inclusion of the $O(\alpha_s^2)$ terms in the spectral density 
would lead to a corresponding modification of the effective continuum threshold, 
with the net effect upon the form factor of only a few percent.] 

\item
{\it The model for the effective continuum threshold}: This very quantity determines 
to a great extent the value of the form factor extracted from the sum rule. 
The possibility to fix this threshold is the weak point of 
the approaches based on sum rules, which limits their predictivity \cite{lms_svz}. 

We use the same universal effective continuum threshold in two- and three-point sum rules.  
This allows us to relate the value of the threshold to the 
pion decay constant, known experimentally. 
We therefore have no free numerical parameters in our analysis. 

There are at least two arguments in favour of our choice of $s_0(Q^2)$: 

First, we have demonstrated that it leads to the correct asymptotic behaviour of the 
pion form factor at $Q^2\to\infty$. 

Second, we expect our approach to work better with the increase of $Q^2$. We have seen that 
it works very well already at relatively small $Q^2=1$--$4$ GeV$^2$ (recall that we have no 
numerical parameters to be 
tuned to reproduce the data). Therefore, we {believe} that for all $Q^2>1$ GeV$^2$ we give 
reasonable predictions. 
However, we cannot control the accuracy of our predictions for the form factor and cannot provide 
any error estimates. 
\item
We can, however, control much better the relative weights of the $O(1)$ and $O(\alpha_s)$ contributions to 
the form factor: their ratio is practically independent of the model for the continuum threshold and is 
determined to great extent by the corresponding $O(1)$ and $O(\alpha_s)$ double spectral densities. 
Here, our results convincingly show that the $O(\alpha_s)$ contribution to the pion form 
factor stays at a level below 50\% at 
$Q^2\le 20$ GeV$^2$ and demonstrate in a largely model-independent way 
that the pion form factor is mainly of nonperturbative origin up to very high $Q^2$. 
Thus, our results definitely speak against the pQCD approach to form factors 
at intermediate $Q^2$, referred to as Scenario A in the Introduction, and confirm Scenario B.  
Although obtained without any reference to the shape of the pion DA, our results indirectly 
restrict the pion DA at low values of the renormalization scale: 
For instance, convex DAs of the type of \cite{anis}, close to the asymptotic one, provide the form factor 
compatible with the results reported here. Also a broader class of the DAs, such as, e.g., 
a double-humped DA with a suppressed end-point region of \cite{bakulev2} seems to lead to the pion 
form factor in agreement with our results.
For a conclusive clarification of this point, the analysis of both the 
$O(1)$ and $O(\alpha_s)$ contributions corresponding to this DA is necessary.  
\end{itemize}
Finally, let us notice that the local-duality version, formulated and developed 
by Radyushkin and co-workers, in many cases has definite advantages compared to other versions 
of QCD sum rules: For instance, the standard three-point sum rules cannot go to large $Q^2$ because of polynomial terms, 
the results from light-cone sum rules depend on the light-cone distribution amplitudes. 
Of course, as we have already mentioned above, the numerical results for the form factor from the local-duality 
sum rules depend crucially on the model of the effective continuum threshold used for the calculations, 
but this shortcoming is shared by all versions of QCD sum rules \cite{lms_svz}. 
In addition to this uncertainty, other versions of sum rules have uncertainties 
related to parameters not precisely known, such as the condensates and the distribution amplitudes. 
We therefore believe to provide the most complete analysis of the pion form factor available 
for the time being. 

\acknowledgments 
We would like to thank Alexander Bakulev and Silvano Simula for interesting discussions on the subject. 
D.~M.~was supported by the Austrian Science Fund (FWF) under projects 
P17692 and P20573, by RFBR project 07-02-00551, and by the Alexander von Humboldt-Stiftung.  
V.~B.~was supported by RFBR project 07-02-00417, CRDF grant Y3-P-11-05, 
and president grant MK-2996.2007.2.



\begin{thebibliography}{30}
\bibitem{pqcd}
V.~L.~Chernyak and A.~R.~Zhitnitsky, JETP~Lett.~{\bf 25}, 510 (1977); 
Sov. J. Nucl. Phys. {\bf 31}, 544 (1980);  
G.~P.~Lepage and S.~J.~Brodsky, Phys.~Lett.~{\bf B87}, 359 (1979);  
A.~V.~Efremov and A.~V.~Radyushkin, Theor. Math. Phys. {\bf 42}, 97 (1980); 
Phys.~Lett.~{\bf B94}, 245 (1980). 
\bibitem{jackson}
G.~R.~Farrar and D.~R.~Jackson, Phys.~Rev.~Lett.~{\bf 43}, 246 (1979).  
\bibitem{cz}V.~L.~Chernyak and A.~Zhitnitsky, 
Phys.~Rep.~{\bf 112}, 492 (1982).
\bibitem{sterman} 
H.-N.~Li and G.~Sterman, Nucl.~Phys.~{\bf B381}, 129 (1992). 
\bibitem{isgur}
N.~Isgur and C.~H.~Llewellyn Smith, Phys.~Lett.~{\bf B217}, 535 (1989).
\bibitem{nesterenko} V.~A.~Nesterenko and A.~V.~Radyushkin, Phys.~Lett.~{\bf B115}, 410 (1982).
\bibitem{ioffe}
B.~L.~Ioffe and A.~V.~Smilga, Phys.~Lett.~{\bf B114}, 353 (1982).
\bibitem{bakulev}A.~P.~Bakulev and A.~V.~Radyushkin, Phys.~Lett.~{\bf B271}, 223 (1991). 
\bibitem{braun1}V.~M.~Braun and I.~Halperin, Phys. Lett. {\bf B328}, 457 (1994). 
\bibitem{simula} F.~Cardarelli et al., Phys.~Lett.~{\bf B332}, 1 (1994).
\bibitem{m}
D.~Melikhov, Phys.~Rev.~{\bf D53}, 2460 (1996); Eur. Phys. J. direct {\bf C4}, 2 (2002) [hep-ph/0110087]. 
\bibitem{anis} V.~V.~Anisovich, D.~I.~Melikhov, and V.~A.~Nikonov, 
Phys.~Rev.~{\bf D52}, 5295 (1995); Phys.~Rev.~{\bf D55}, 2918 (1997). 
\bibitem{braun2}V.~M.~Braun, A.~Khodjamirian, and M.~Maul, Phys. Rev. {\bf D61}, 073004 (2000). 
\bibitem{braguta2}V.~V.~Braguta and A.~I.~Onishchenko, Phys.~Lett.~{\bf B591}, 255 (2004). 
\bibitem{braguta}V.~V.~Braguta and A.~I.~Onishchenko, Phys.~Lett.~{\bf B591}, 267 (2004). 
\bibitem{svz}
M.~A.~Shifman, A.~I.~Vainshtein, and V.~I.~Zakharov,
Nucl.\ Phys.\ {\bf B147}, 385 (1979).
\bibitem{ms}D.~Melikhov and S.~Simula,
Eur.~ Phys.~J. {\bf C37}, 437 (2004).
\bibitem{colangelo}
M.~Neubert, Phys.~Rev.~{\bf D47}, 4063 (1993); \\
P.~Colangelo, F.~De~Fazio, and N.~Paver, Phys.~Rev.~{\bf D58}, 116005 (1995).
\bibitem{radpol}A.~V.~Radyushkin, Acta~Phys.~Polon.~{\bf B26}, 2067 (1995).
\bibitem{lucha}W.~Lucha and D.~Melikhov,
Phys.~Rev.~{\bf D73}, 054009 (2006); 
Phys.~Atom.~Nucl.~{\bf 70}, 891 (2007). 
\bibitem{radsmallq2}V.~A.~Nesterenko and A.~V.~Radyushkin, 
Pisma. Zh.~Exp.~Theor.~Phys~{\bf 39}, 576 (1984); \\
I.~I.~Balitsky and A.~V.~Yung, Phys.~Lett.~{\bf B129}, 328 (1983). 
\bibitem{bakulev2}A.~P.~Bakulev et al., Phys.~Rev.~{\bf D70}, 033014 (2004).
\bibitem{tiktopoulos}J.~Cornwall and G.~Tiktopoulos, Phys.~Rev.~{\bf D13}, 3370 (1976). 
\bibitem{lms_lc}
W.~Lucha, D.~Melikhov, and S.~Simula, Phys.~Rev.~{\bf D75}, 096002 (2007).  
\bibitem{data_largeQ2}C.~J.~Bebek, Phys.\ Rev.\ {\bf D17}, 1693 (1978).
\bibitem{craig}P.~Maris and C.~D.~Roberts, Phys.~Rev.~{\bf C58}, 3659 (1998).  
\bibitem{lms_svz}W.~Lucha, D.~Melikhov, and S.~Simula,
Phys.~Rev.~{\bf D76}, 036002 (2007);   
Phys.~Lett.~{\bf B657}, 148 (2007). 
\end{thebibliography}
\end{document}